\documentclass[fleqn,usenatbib]{mnras}

\usepackage{newtxtext,newtxmath}

\usepackage[T1]{fontenc}
\usepackage{ae,aecompl}

\usepackage{graphicx}	
\usepackage{amsmath}	
\usepackage{amssymb}	

\newcommand{\angstrom}{\mbox{\normalfont\AA}}

\title[Are the Carriers of DIBs and ERE the Same?]{Are the Carriers of Diffuse Interstellar Bands and Extended Red Emission the same?}

\author[Thomas S. -Y. Lai et al.]{
Thomas S. -Y. Lai$^{1}$\thanks{E-mail: shaoyu.lai@utoledo.edu},
Adolf N.Witt$^{1}$,
Carlos Alvarez$^{2}$,
Jan Cami$^{3,4,5}$
\\
$^{1}$Ritter Astrophysical Research Center, University of Toledo, Toledo, OH 43606, USA\\
$^{2}$W. M. Keck Observatory, 65-1120 Mamalahoa Hwy., Kamuela, HI 96743-8431, USA\\
$^{3}$Department of Physics and Astronomy, The University of Western Ontario, London, ON N6A 3K7, Canada\\
$^{4}$Institute for Earth and Space Exploration, The University of Western Ontario, London, ON N6A 3K7, Canada\\
$^{5}$SETI Institute, 189 Bernardo Avenue, Suite 100, Mountain View, CA 94043, USA
}

\date{Accepted 2020 January 13. Received 2019 December 24; in original form 2019 July 8.}

\pubyear{2020}

\begin{document}
\label{firstpage}
\pagerange{\pageref{firstpage}--\pageref{lastpage}}
\maketitle

\begin{abstract}
We report the first spectroscopic observations of a background star seen through the region between the ionization front and the dissociation front of the nebula IC 63. This photodissociation region (PDR) exhibits intense extended red emission (ERE) attributed to fluorescence by large molecules/ions. We detected strong diffuse interstellar bands (DIB) in the stellar spectrum, including an exceptionally strong and broad DIB at $\lambda$4428. The detection of strong DIBs in association with ERE could be consistent with the suggestion that the carriers of DIBs and ERE are identical. The likely ERE process is recurrent fluorescence, enabled by inverse internal conversions from highly excited vibrational levels of the ground state to low-lying electronic states with subsequent transitions to ground. This provides a path to rapid radiative cooling for molecules/molecular ions, greatly enhancing their ability to survive in a strongly irradiated environment. The ratio of the equivalent widths (EW) of DIBs $\lambda$5797 and $\lambda$5780 in IC 63 is the same as that observed in the low-density interstellar medium with UV interstellar radiation fields (ISRF) weaker by at least two orders of magnitude. This falsifies suggestions that the ratio of these two DIBs can serve as a measure of the UV strength of the ISRF. Observations of the nebular spectrum of the PDR of IC 63 at locations immediately adjacent to where DIBs were detected failed to reveal any presence of sharp emission features seen in the spectrum of the Red Rectangle nebula. This casts doubts upon proposals that the carriers of these features are the same as those of DIBs seen at slightly shorter wavelengths.

\end{abstract}



%
%
%
\begin{figure*}
 	\includegraphics[width=0.75\textwidth]{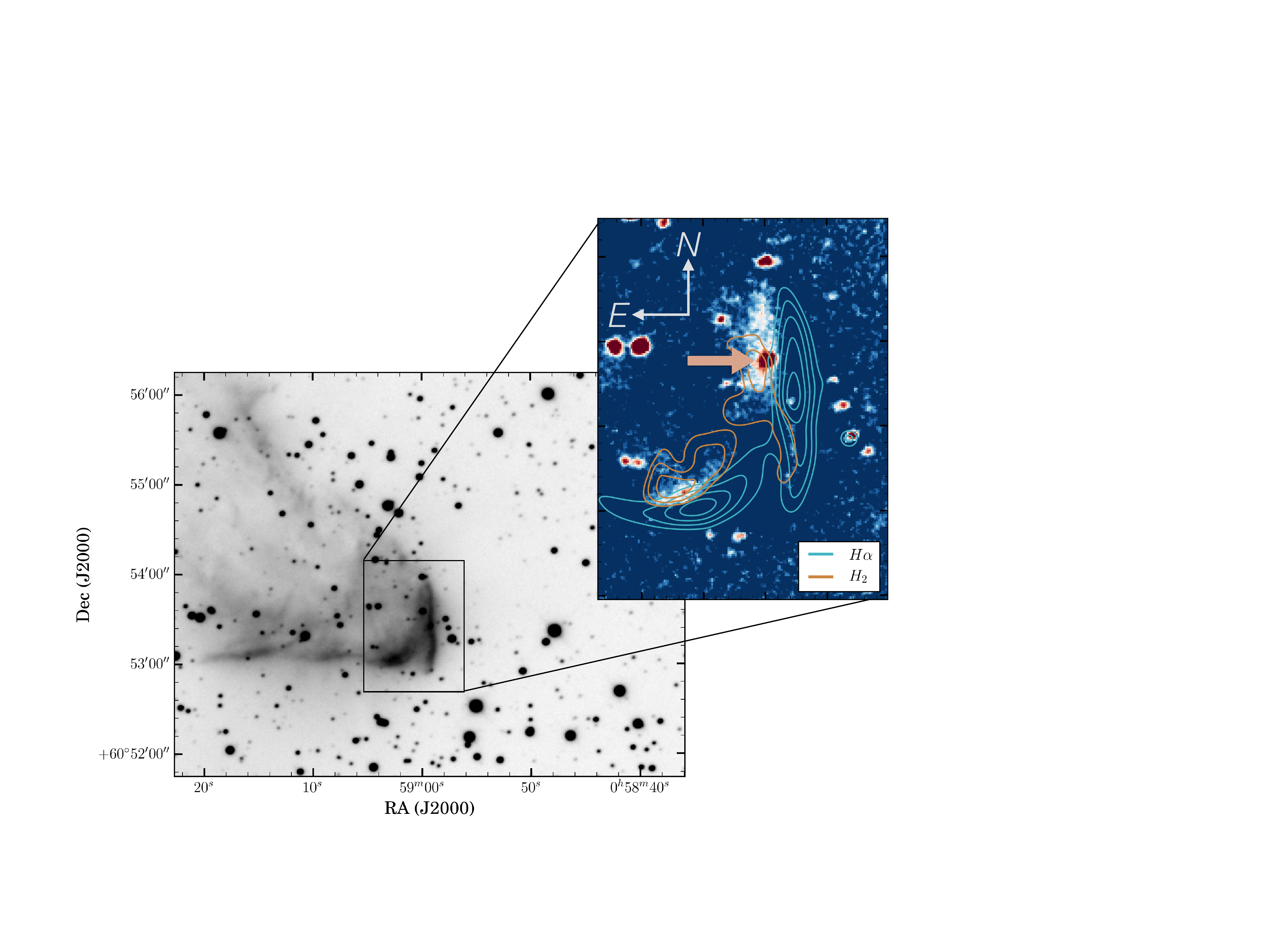}
    \caption{The morphology of IC63. The images were taken with the 20-inch RC Optical Systems reflector of the Rancho Del Sol Observatory in Camino, CA. The gray scale image was taken with the wide-band Astrodon E-Series red band filter. The ERE map in the inset, with the size of 90$\arcsec$ $\times$ 72$\arcsec$, was created by digital division of the red image (corrected for the H$\mathrm{\alpha}$ and [S$\mathrm{_{II}}$] emission) and the green image. Two of the brightest ERE filaments in IC 63 are shown in lighter color. Star \#46, denoted by the arrow, appears right behind the strongest ERE emission in the PDR, produced by radiation from the early B-star $\gamma$ Cas, incoming from the SW direction. The peak intensities of ERE are found immediately behind the hydrogen ionization fronts in regions also occupied by UV-pumped rotationally excited molecular hydrogen (for detail, see \citet{Lai2017}).}
    \label{fig:star46_ERE}
\end{figure*}

\section{Introduction}
\subsection{Background}
The diffuse interstellar bands (DIBs) and extended red emission (ERE) share a common mystery: the carriers of the two interstellar phenomena are largely unidentified despite the intensive research that has been conducted throughout their history, spanning many decades. DIBs are a set of ubiquitous absorption features that have been observed in the optical and near-infrared spectra of stars that lie behind diffuse interstellar clouds \citep[see][for reviews]{Herbig1995, Sarre2006, Snow2014, Krelowski2018}. \citet{Heger1922} first discovered two diffuse absorption features at 5780 $\angstrom$ and 5797 $\angstrom$, followed by a more systematic study of DIBs and a recognition of their interstellar origin by \citet{Merrill1938}. \hspace{3cm} By now, over 500 DIBs have been cataloged, with equivalent widths ranging from a few m$\angstrom$ to a few 1000 m$\angstrom$ \citep{Hobbs2008, Hobbs2009, Fan2019}, with estimated column densities of individual DIB carriers for standard lines of sight of the order $\mathrm{N} \sim 10^{13}$cm$^{-2}$ \citep{Herbig1993}. The lack of strong correlations among about 10\% of the strongest  DIBs suggests that most DIBs are the product of different, distinct carriers. To identify the carriers of DIBs is difficult; only C$_{60}^{+}$ has been identified as the carrier of two or more near-infrared DIBs (\citealt{Campbell2015}; \citealt{Cordiner2019}; \citealt{Lykhin2019}, but also see \citealt{Galazutdinov2017}). Nevertheless, the DIB carriers as a whole must share a common characteristic: stability, which enables them to survive under the harsh UV irradiation conditions in the diffuse interstellar medium (ISM). The most likely carriers of DIBs are thought to be carbonaceous molecules and molecular ions, including polycyclic aromatic hydrocarbons \citep[PAHs;][]{Crawford1985, Leger1985, vanderZwet1985, Salama1996}, linear carbon chain molecules \citep{Douglas1977, Oka2013}, fullerenes of various sizes \citep{Kroto1985, Omont2016}, fulleranes \citep{Webster1992}, and polyacenes \citep{Omont2019}. 

To aim for an exhaustive identification of the exact carriers of DIBs is nearly impossible, because even for a molecule with 6 heavy atoms there exist over 1000 different possible structures; as for a molecule with $\sim$30 atoms, this number can easily exceed a million \citep{Jones2016}. Thus, the truly remarkable fact about the number of detected DIBs is that there are so few (only $\sim$ 500) rather than many more. Thus, it appears reasonable to argue that a great portion of the potential carriers of DIBs are unlikely to survive under interstellar conditions, and their stability/instability against photo-dissociation is most likely the discriminator as to whether small molecular structures can persist under interstellar condition for long enough to build up sufficiently large column densities.

Another long-standing question which remains open is the emission mechanism and corresponding carriers of the ERE, a pervasive photon-driven luminescence process observed in a wide variety of interstellar environments \citep{Witt2004}. While ERE has been observed in many reflection nebulae \citep{Witt1990}, it is absent in reflection nebulae with illuminating stars that have T$\mathrm{_{eff}}$ < 10$^4$ K \citep{Darbon1999}, suggesting that ERE is excited only by far-ultraviolet (FUV) photons, which are absent in the flux from stars with a lower temperature. Subsequent studies have shown that only photons with energies E $\geq$ 10.5 eV contribute to the excitation of ERE \citep{Witt2006, Lai2017}, confirming the suggestion implied by the findings of \citet{Darbon1999}. Among a variety of proposed ERE models \citep{Witt2014}, recurrent fluorescence, also known as Poincar\'e fluorescence \citep{Leger1988}, in highly isolated molecules or molecular ions explains the production mechanism of the ERE fairly well by meeting all the observational constraints \citep{Lai2017}, in particular the high photon conversion efficiency found in the diffuse ISM \citep{Gordon1998} and the large ($\geq$ 8 eV) difference between the excitation and emission energies. 

Recurrent fluorescence occurs in molecules or molecular ions that are highly isolated, conditions expected in the interstellar medium where collision timescales are many orders of magnitude longer than timescales for internal molecular processes. When such a carrier absorbs a FUV photon, its energy undergoes rapid internal conversion (IC), transferring the electronic excitation energy into vibrational energy of the electronic ground state. For a marginally stable particle, several modes of relaxation are then possible:
(1) ejection of particles (electrons, hydrogen atoms, molecular fragments), ultimately leading to destruction; (2) cooling via mid-infrared vibrational transitions on a fairly long timescale of seconds; and (3) inverse internal conversions (IIC; \citealt{Leger1988}), followed by emission of one or more low-energy optical photons. The latter process can occur on a much more rapid timescale ($\simeq 2 \times 10^{-5}$s ; \citealt{Chandrasekaran2014}) than the slow emission of mid-infrared vibrational photons ($\simeq 3$s), but it requires a highly vibrationally excited ground state and the presence of a low-lying electronic state with excitation energy E $\lesssim$ 2.3 eV above ground and a high transition probability to the ground state. The detection of red photons resulting from recurrent fluorescence in C$_{6}^{-}$ ions was recently reported by \citet{Ebara2016}, confirming experimentally the viability of this process. More recently, recurrent fluorescence photons from C$_{4} ^{-}$ have also been demonstrated experimentally \citep{Yoshida2017}.

Depending on the energy available from the initial excitation, recurrent fluorescence can result in the emission of multiple optical photons from a single excitation, leading potentially to high fluorescence efficiency of several hundred percent. This could explain the high photon conversion efficiency observed for the ERE \citep{Gordon1998}. Molecules meeting this condition are thus able to shed a major fraction of their vibrational energy very rapidly, in particular when compared to the much slower cooling via mid-infrared vibrational emission, which then greatly reduces the likelihood of photo-fragmentation. In recent years, experimental evidence supporting the role of recurrent fluorescence in rapidly cooling highly excited molecular ions and small clusters has accumulated \citep{Ferrari2018, Hansen2017, Bernard2017}. As a result, small molecules and molecular ions in the interstellar medium meeting these conditions are more stable compared to similarly sized structures lacking this capacity. Consequently, their relative abundances will be increased substantially to column densities where these same molecules in a cold state can produce observable low-energy optical absorption features. It appears possible, therefore, that many DIBs observed in the red part of the optical spectrum are produced by absorption by the same carriers in a cold state that are responsible for the ERE by emission when they are in a vibrationally hot state.

\subsection{DIB/ERE Connection}
\label{DIB_ERE_connection}
A suggestion of a possible connection between DIBs and ERE based on these ideas was made recently \citep{Witt2014}. Both DIBs and ERE are produced with highest efficiency in the diffuse ISM, preferring atomic hydrogen environments and appearing with lower efficiency in high-UV environments, where carriers are more likely subject to photodissociation \citep{Smith2002}. Both phenomena exhibit non-unique global spectra \citep{Witt1990, Smith2002, Hobbs2009}, which vary from one object to another or from one line of sight to another. This suggests that the likely carriers of DIBs and ERE in each case are large families of molecules and/or molecular ions, which remain marginally stable in the diffuse ISM, with abundances that are sensitive to local environmental conditions. For both DIB and ERE processes, carriers involve electronic transitions with energies predominantly in the range of 1.4--2.3 eV above the ground state \citep{Witt2014}. If one compares the histogram of the DIB density per unit wavelength with the typical ERE spectrum, they look remarkably similar, rising steeply longward of 5400 $\angstrom$ and peaking within the range of 6000--7000 $\angstrom$ (see Fig. 1 in \citealt{Witt2014}). For these reasons it appears legitimate to ask the question whether the carriers of DIBs and ERE are the same or at least members of closely related families of molecular particles. Thus, the main goal of this paper is to carry out an observational investigation of the possible relation between DIB and ERE carriers. 

To assess such a possible relationship, we focus on the detection of $\lambda$4428, since it is the broadest and strongest feature among all the observed DIBs, with a FWHM $\approx$ 20 $\angstrom$ and a typical EW of a few $\angstrom$. It was first reported by \citet{Beal1937}, followed by the first profile measurement provided by \citet{Baker1949}. A more comprehensive study of the $\lambda$4428 profile was conducted by \citet{Snow2002b}, who pointed out that the intrinsic profile is best fitted by a Lorentzian. 

\subsection{IC 63 and Star \#46}
Establishing a direct correlation between the two phenomena is difficult since DIBs are usually observed in the line of sight towards a distant star, while ERE is observed most easily in a diffuse emission environment adjacent to a hot illuminating star. Fortunately, the reflection nebula IC 63 gives us an opportunity to observe the two phenomena simultaneously in the same line of sight. The existence of ERE in IC 63 was first reported by \citet{Witt1990} and recently  verified by \citet{Lai2017}. As shown in Fig. \ref{fig:star46_ERE}, the region with the most intense ERE in IC 63 coincides with the line of sight to a faint background star \#46 of spectral type A5V \footnote{It is originally labelled as an A7V star. See Sec. \ref{data reduction} for detail.} \citep{Andersson2013}, whose spectrum is sufficiently free of photospheric absorption lines to permit the observation of some of the stronger DIBs \citep{Herbig1995}. The amount of reddening of star \#46 is in close agreement with that expected from the passage of its light through the PDR in IC 63, where the most intense ERE is seen \citep{Habart2004, France2005}. Thus, we can examine the same interstellar material responsible for the ERE in IC 63 for the presence of DIBs. In other words, we can test the hypothesis that DIBs and ERE are produced by the same carriers in the sense that the absence of DIBs would falsify our hypothesis of common carriers for the two processes. If DIBs are present in such a line of sight, it could either be consistent with the hypothesis or at least be an indication that the two phenomena simply coexist but are independent of each other. We note that extensive studies of DIBs in nebular environments have found DIBs to be either weak relative to the observed reddening of the stars or absent altogether (\citealt{Snow1995}, and references therein). Thus, observing strong DIBs along the line of sight through the ERE filament in IC 63 would be rather exceptional and strongly supportive of a connection between the carriers of the two phenomena.

\subsection{RR Emission Features \& the Probe of ISRF}
A second test made possible by the fortuitous alignment of an ERE filament in the reflection nebula IC 63 with a suitable background star is an inquiry into the nature of the unidentified, sharp optical emission features seen in the spectrum of the Red Rectangle nebula (RR-features) and the R Coronae Borealis star V854 Centauri \citep{Rao1993, Oostrum2018} and their possible relation to DIBs \citep{Scarrott1992, Sarre1995, Glinski2002}. The close proximity in wavelengths of some of the RR-features to the wavelengths of prominent DIBs has led to the suggestion that the RR-features are DIBs seen in emission \citep{Sarre1995, vanWinckel2002, Adams2019}, and thus produced by the same carriers. One argument against this suggestion comes from the fact that no corresponding DIB absorption features were observed in the spectrum of the deeply embedded central star of the Red Rectangle, HD 44179, indicating an absence of the DIB carriers from the line of sight to the star. We note, however, that the RR-features are observed in the Red Rectangle only at lines of sight that are offset from the direction to the central star.

In the Red Rectangle nebula, the RR-features show a close spatial correlation with the brightest ERE structures \citep{Schmidt1991}. A long-slit observation of star \#46 (see \textsection \ref{DCT_obs}) through the region of the most intense ERE in IC 63 will also yield a spectrum of the nebula with the potential of detecting RR-features, if present. Thus, a test of the possible correlation between DIBs and corresponding RR-features seen along the same line of sight is possible for the first time. 

Finally, the exceptionally high UV radiation field in IC 63 (G$_{0} \gtrsim $150; \citealt{Andrews2018}) allows us to investigate the ratio of two DIBs, $\lambda5797$ and $\lambda5780$, which has been suggested to be a useful measure of the local interstellar radiation intensity \citep{Cami1997, Vos2011}. If true, the ratio of EW5797/EW5780 should deviate significantly from that found in the diffuse ISM with typically low interstellar radiation field densities.

The layout of the paper is as follows: In \textsection \ref{sec:observation} we detail the DeVeny and DEIMOS observations and the data reduction process; In \textsection \ref{results} we present the results of our observations, including the photometry of star \#46, measurements of $\lambda$4428 as well as other DIBs, and the non-detection of the RR features, followed by a discussion of these topics in \textsection \ref{discussion}. \textsection \ref{summary} contains the summary and conclusions of the paper.

\section{Observation and data reduction}
\label{sec:observation}
We recorded spectra of star \#46, seen through the line of sight of the brightest ERE filament in the reflection nebula IC 63, by using the DeVeny spectrograph on the Discovery Channel Telescope (DCT) and the DEep Imaging Multi-Object Spectrograph \citep[DEIMOS;][]{Faber2003} on the Keck II telescope.

\begin{table*}
	\centering
	\caption{DCT/DeVeny Observation summary}
	\label{Tab:observation}
	\begin{tabular}{lcccccc} 
		\hline\hline
		Target & RA  & DEC  & V & Spectral Type & E$_{\mathrm{B-V}}$ & Obs. time \\[0.1mm]
 		& (J2000) & (J2000) & (mag.) & & (mag.) & (s) \\
        \hline
		IC 63 \#46 & 00:58:59.9 & 60:53:35.2 & 18.24 & A5 V & 1.23 & 24300\\
		Feige 11 & 01:04:21.6 & 04:13:37.0 & 12.07 & sdB2 VIHe4 & 0 & 320\\
		HD 183143 & 19:27:26.6 & 18:17:45.2 & 6.86 & B6 Ia & 1.27 & 10\\
		HD 194839 & 20:26:21.5 & 41:22:45.6 & 7.49 & B0.5 Iae & 1.18 & 40\\
		HD 229059 & 20:21:15.4 & 37:24:31.1 & 8.70 & B2 Iabe & 1.71 & 150\\ 
		\hline
	\end{tabular}
\end{table*}

\subsection{DCT/DeVeny observation}
\label{DCT_obs}
We used the DeVeny spectrograph with a 300 g/mm grating (R = 920), blazed at 4000 $\angstrom$ (1$\mathrm{^{st}}$-order), to observe the spectrum of star \#46 as well as the spectra of the flux/spectral standard stars. Our DCT result was based on three nights of observations: 9/29/2016, 9/30/2016, and 10/10/2017. The spectrum of star \#46 was obtained by combining 27 images with a total of 24300 s (405 min) of exposure. The wavelength coverage extends over the range of $\sim$ 4000 -- 6000 $\angstrom$. The long-wavelength limit was set by the fact that the spectrum produced by the DeVeny spectrograph is subject to $\mathrm{2^{nd}}$-order leakage above 6000 $\angstrom$. In order to perform an accurate sky subtraction for star \#46, the placement of the slit, $1\arcsec \times 1.9\arcmin$, was centered at the position of \#46 star (RA=$00^{h} 58^{m} 59.9^{s}$, Dec=$60^{\circ} 53' 35.2''$) and aligned in the north-south direction, following the morphology of the nebula's ERE filament. This allowed us to avoid contamination from the photoionization front of the nebulosity, a few arcsecond west of star \#46, where strong H$\alpha$ emission may saturate the image. For flux calibration, we used Feige 11 from \citet{Oke1990} as the flux standard star. In order to confirm the detection of $\lambda$4428, we employed the same instrumental configuration used for observing star \#46 to observe three other stars that were known to have prominent $\lambda$4428 and were accessible during the observing period: HD 183143, HD 194839, and HD 229059. The details of these observation are summarized in Table \ref{Tab:observation}.

\subsection{Keck-II/DEIMOS observation}
In addition to the DCT observation, we acquired Keck-II/DEIMOS R$\sim$1600 spectra of star \#46 using the 900ZD grating, blazed at 5500 $\angstrom$ (1$\mathrm{^{st}}$-order), with the $1 \arcsec$ wide slit, covering a wavelength range $\sim$4000--7000 $\angstrom$. The wavelength coverage from DEIMOS provides a valuable supplement to the DeVeny spectrum, particularly for wavelengths above 6000 $\angstrom$. The somewhat higher spectral resolution also allowed us to search for the sharp RR-emission features in the adjacent nebular spectrum in this wavelength range. The slit was aligned in the north-south direction and centered at star \#46, similar to our observation with the DCT (see Sec. \ref{DCT_obs}). By stacking 9 exposures, the total exposure time of star \#46 was 10200 s (170 min). Our DEIMOS data was based on a half night observation on 7/16/2017.

\subsection{Ancillary data}
\label{ancillary data}
The University of Chicago (UoC) DIB database contains a large number of echelle spectra from the Apache Point Observatory Diffuse Interstellar Band (DIB) collaboration, with a resolution of R=38,000 and a S/N$\sim$1000 at 6400$\angstrom$. This database is publicly available online \footnote{http://dib.uchicago.edu/public/index.html}. In particular, we used the spectrum of the unreddened A5V star HD194837 (E$_{\mathrm{B-V}}$ = 0.01) from this database as a spectral standard and compared it with star \#46 to set the continuum for the renormalization (see Sec. \ref{data reduction} for details). We also used the measured values of EW4428 and E$_{\mathrm{B-V}}$ for a large sample of stars from this database for a comparison with our results.

\subsection{Data reduction}
\label{data reduction}
The long-slit advantage provided by the sightline through IC 63 to star \#46 offers the possibility to observe both the DIB absorption features in the stellar spectrum and the potential emission features from the adjacent nebular spectrum excited by the nearby early B-star, $\gamma$ Cas, as well as the night sky spectrum seen in lines of sight immediately adjacent to that of star \#46 and the ERE filament of IC 63. An accurate sky subtraction is a prerequisite for extracting spectra for both the star and nebula. We used $\textit{IRAF}$ for the image pre-processing and \textit{python} for the 1-D spectrum extraction. The extraction widths in individual exposures for star \#46 from the DeVeny and DEIMOS spectrographs are 2.72$\arcsec$--3.06$\arcsec$ and 2.37$\arcsec$--3.55$\arcsec$, respectively. The differences of the extraction widths are due to the change in seeing condition of each pointing. Once the 1-D spectrum was extracted, we excluded the background by carefully subtracting the spectrum of nebula and sky adjacent to the star's position. For star \#46, we measured the synthetic photometric brightness as B=19.63 and V=18.24 by convolving the reduced stellar spectrum with the B and V band passes and integrating over wavelengths, with calibration provided through the identical process applied to the spectrum of the flux standard, Feige 11. As for the nebular spectrum, we used an extraction width of 7.82$\arcsec$ to extract the nebular emission that is $\sim$11$\arcsec$ south of star \#46 and employed the sky spectrum in off-nebula portions for the sky correction.

\begin{figure}
 	\includegraphics[width=1.0\columnwidth]{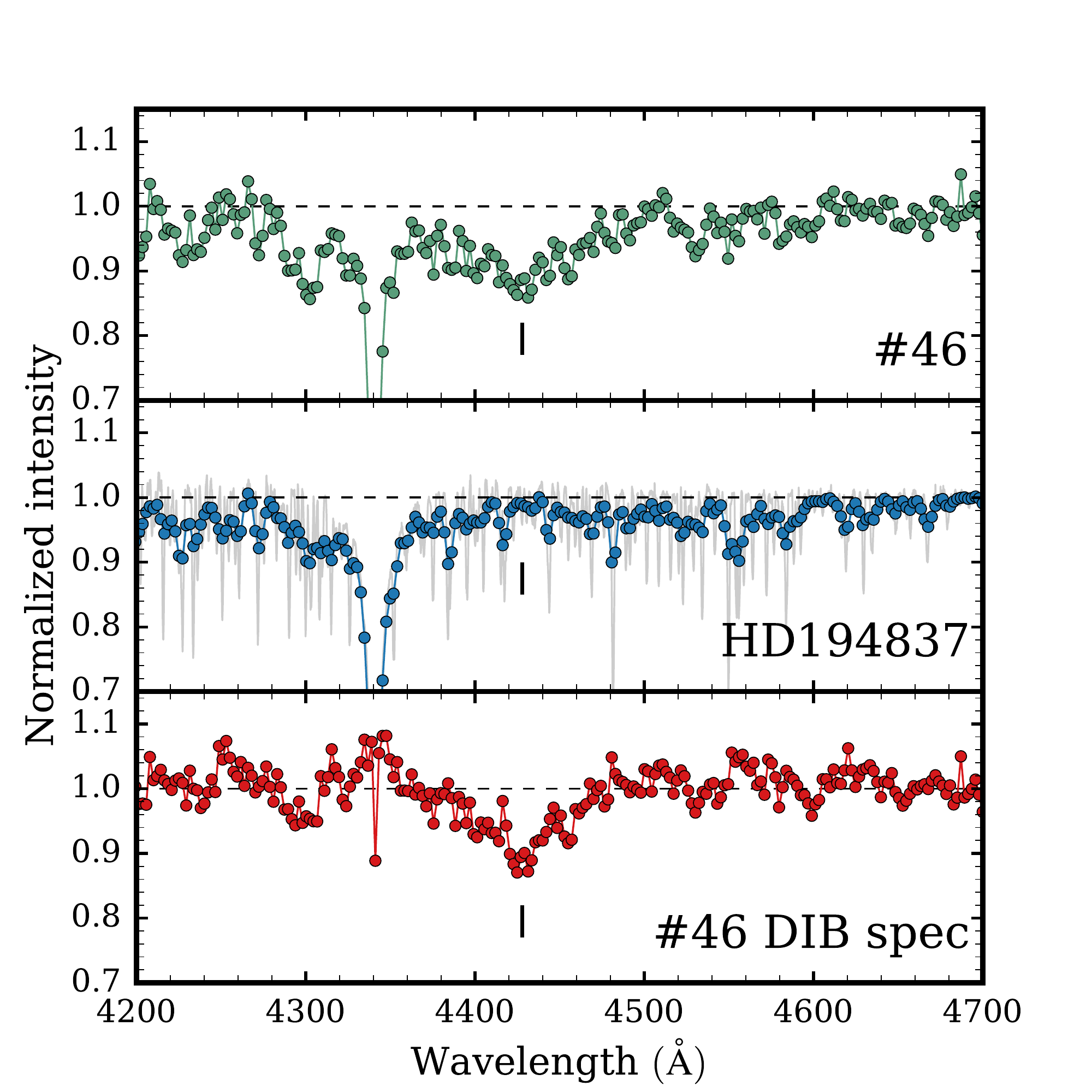}
    \caption{Star \#46 DIB spectrum centered at $\lambda$4428 (bottom) was produced by dividing the original star \#46 spectrum (top; taken with DCT/DeVeny) by the degraded HD194837 spectrum (middle), to avoid the confusion caused by the strong H$_{\gamma}$ absorption line near 4340 $\angstrom$. The original spectrum of HD194837 retrieved from the UoC database is shown in light gray colour. The vertical bar in each panel indicates the wavelength of 4428 $\angstrom$.}
    \label{fig:star46_norm}
\end{figure}

The observed DIBs have a range in widths. It is relatively easy to detect narrow and deep absorption features, but hard to detect broad, shallow DIBs due to the difficulty in setting the continuum over such features. Among the set of very broad DIBs, $\lambda$4428 is the strongest, with FHWM $\sim$ 20 $\angstrom$, situated in a region contaminated by stellar photospheric lines. The extended wing of the strong H$\gamma$ absorption line in A stars makes determining the continuum on the short-wavelength side of $\lambda$4428 even more challenging. In addition, the entire blue spectrum of A5V stars is crowded with numerous weak, narrow absorption lines of various strengths. At the resolution of the DeVeny spectrograph (R = 920; $\Delta \lambda \sim$ 4.8 $\angstrom$), these lines are unresolved, appearing as a wavy quasi-continuum. This causes the profile of the $\lambda$4428 band to acquire corresponding distortions, which become particularly serious in the extended wings of the band. This can be corrected by dividing the spectrum of star \#46 by a spectrum of an unreddened star of identical spectral type with identical spectral resolution, which should remove these distortions to a fair degree. This method was used with good success by \citet{Yuan2012}. We chose the high resolution spectrum of HD 194837, an unreddened A5V star, from the UoC DIB database and degraded it to match the resolution of the star \#46 spectrum (R = 920) before performing this division. The final DIB spectrum of star \#46 centered at 4428 $\angstrom$ is shown in Fig. \ref{fig:star46_norm}. Note that the spectral type of star \#46 was classified as A7V in \citet{Andersson2013}, but we found that an A7V star has a shallower H$\gamma$ absorption feature than is observed in star \#46, which makes the attempt to exclude H$\gamma$ by dividing both spectra unsatisfactory. By contrast, we found an A5V stellar spectrum to provide a better match.

\begin{figure}
 	\includegraphics[width=1.0\columnwidth]{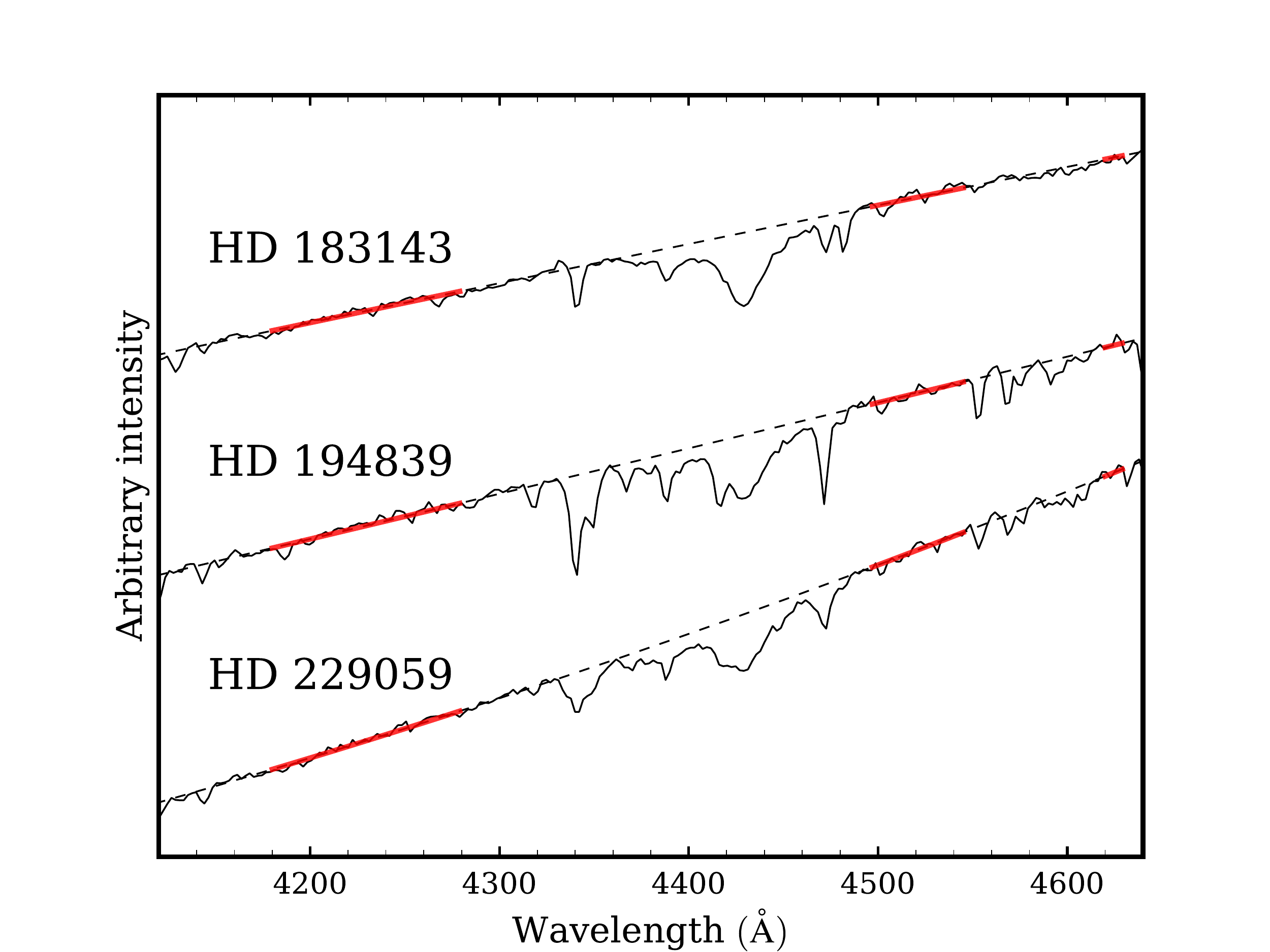}
    \caption{Spectra of the three $\lambda$4428 standard stars before applying normalization. $\lambda$4428 is such a broad feature that it reaches the continuum at $\sim$4360$\angstrom$ and $\sim$4490$\angstrom$. The red lines show the range where the continua are fitted. These spectra were obtained with the DCT/DeVeny system.}
    \label{fig:4428_std_star_rawdata}
\end{figure}

As for the $\lambda$4428 standard stars, which are all of much earlier spectral types,  we established the nearby continua for the renormalization by fitting the continua with a polynomial as shown in Fig. \ref{fig:4428_std_star_rawdata}. Based on Fig. \ref{fig:4428_std_star_rawdata}, we clearly see that the wide $\lambda$4428 feature reaches the continuum at $\sim$4360$\angstrom$ and $\sim$4490$\angstrom$. Thus, we renormalized the spectra by assigning continua outside this wavelength range as shown in red in Fig. \ref{fig:4428_std_star_rawdata}. 

\begin{figure*}
    \includegraphics[width=0.8\textwidth]{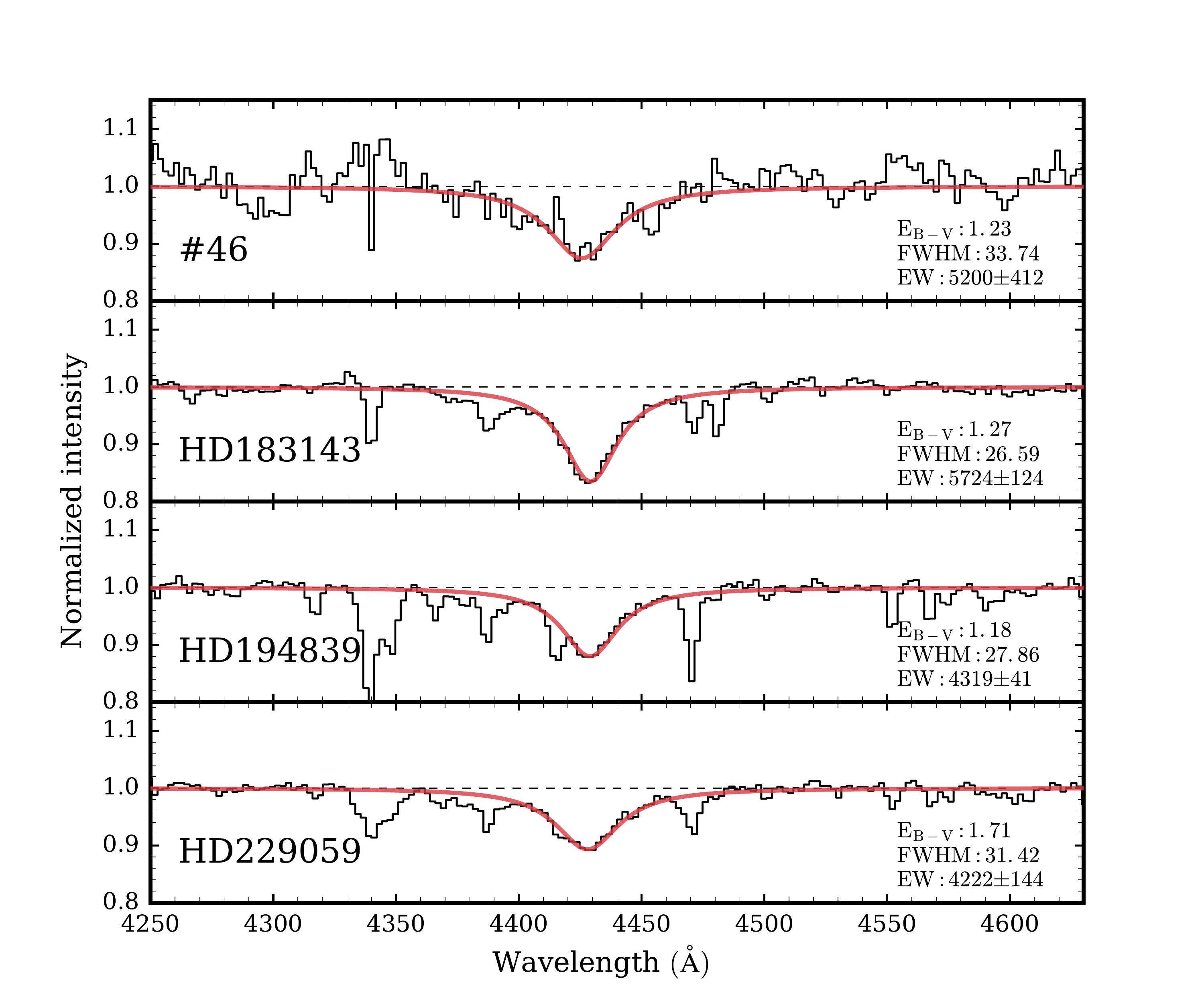}
    \caption{Observed spectra of $\lambda$4428 profiles from star \#46 and three standard stars in our observation. We estimate the $\lambda$4428 FWHM and EW by fitting each spectrum with a Lorentzian profile as shown in red. The units of FWHM and EW are in [$\angstrom$] and [m$\angstrom$], respectively.}
    \label{fig:4428_spec_std}
\end{figure*}

\section{Results}
\label{results}
\subsection{Photometry of star \#46}
There exists only a single photometric measurement of star \#46 from \citet{Andersson2013} with V=16.18 $\pm$ 0.01 and E$_{\mathrm{B-V}}$ = 1.00, with spectral type A7V. In our study, we carried out the photometry of star \#46 by comparing its flux with that of our flux standard star Feige 11 (V = 12.07). This led to a colour excess E$_{\mathrm{B-V}}$ =  1.23 and a magnitude of V = 18.24. The discrepancy with the earlier measurements is consistent with an inappropriate sky background value used in the SDSS photometry pipeline employed by \citet{Andersson2013}. 

It is important to note that the reddening of star \#46 is caused predominantly by the passage of its light through the PDR in IC 63, where most of the ERE is seen. The hydrogen column densities of the PDR in IC 63 has been estimated as N(H$_{2}$) $\sim$ 5 $\times$ 10$^{21}$ cm$^{-2}$ \citep{Jansen1995, Habart2004, France2005}. Applying the gas-to-dust ratio of \citet{Bohlin1978}, N(HI + H$_{2}$) / E$_{\mathrm{B-V}}$ = 5.8 $\times$ 10$^{21}$ cm$^{-2}$, we arrive at an estimated reddening due to the PDR passage of E$_{\mathrm{B-V}}$ $\sim$ 0.95, leaving an extra amount of E$_{\mathrm{B-V}}$ $\sim$ 0.28 to reddening by dust between IC 63 and the location of star \#46. Furthermore, the spectrophotometric distance to star \#46 is about 2.5 kpc. The 3D dust map Bayestar19 of \citet{Green2019} predicts a cumulative reddening of E$_{\mathrm{B-V}}$ $\sim$ 0.43 for the line of sight through star \#46 out to distances of about 3 kpc, suggesting a reddening caused by the PDR in IC 63 of E$_{\mathrm{B-V}}$ $\sim$ 0.80. This provides two fairly consistent estimates that at least 2/3 of the reddening of star \#46 is caused within IC 63, which is significant if a connection between ERE and some DIBs is to be demonstrated.

\subsection{$\lambda$4428 equivalent width and profile}
\label{DIB4428 profile}
To investigate the existence of DIBs towards star \#46, we focused on the detection of the $\lambda$4428 broad feature, given the faintness of the star and the modest spectral resolution provided by the DeVeny spectrograph. The DIB spectra of star \#46 and the three $\lambda$4428 standard stars are shown in Fig. \ref{fig:4428_spec_std}. We estimated the $\lambda$4428 FWHM and EW by fitting each DIB with a Lorentzian profile \citep{Snow2002b}. The uncertainty of the EW caused by the continuum placement was estimated following the description in \citet{Vos2011}: $\sigma _{\mathrm{EW}} \sim \sqrt{2 \Delta \lambda \delta \lambda} / \mathrm{SNR}$, where $\Delta \lambda$ is the integration range and $\delta \lambda$ is the spectral dispersion in $\angstrom$/pix. The SNR was measured using the continuum adjacent to the absorption feature. For star \#46, we found  the EW4428 to be 5200$\pm$412 m$\angstrom$, with the FWHM at 33.74 $\angstrom$. The results for other DIBs detected in the star \#46 spectrum are summarized in Table \ref{Tab:summery_DIBs}.

Unlike other narrower DIBs, detailed studies of the broad DIBs are rare. Only recently,  \citet{Sonnentrucker2018} carried out a homogeneous census on broad DIBs (with FWHM $\geq$ 6$\angstrom$), whose continuum levels were hard to assign when observed with echelle spectrographs, resulting in large uncertainties. Even starting with identical spectra, with different applied methods, one can arrive at a wide range of measured $\lambda$4428 EWs, differing by as much as a factor of 2 \citep{York2014}. In Table \ref{Tab:HD183143_EW4428}, we collected a set of the $\lambda$4428 EWs for HD 183143 from the literature. This comparison shows the EWs reported by different authors vary widely and reflect the complexity of estimating the EW of broad DIBs. The inconsistencies can be attributed mainly to the systematic error resulting from the disagreement in placing the continuum, which tends to be much larger than the statistical errors. 

Even though our EW4428 measurement for $\lambda$4428 standard star HD183143 lies near the upper end of the distribution when compared with other measurements, it is within the uncertainty quoted by \citep{Hobbs2009}. We also note that our placement of the continuum in HD183143 is nearly identical to that of \citet{Herbig1966}, resulting in a central depth of $\lambda$4428 of 16\%. The reason for arriving at a relatively larger EW value is likely due to the larger integration range in our study, which is considerably wider than that applied in some previous works (4400--4460$\angstrom$; see \citealt{Tug1981}, \citealt{Snow2002b}). In addition, in studies prior to \citet{Snow2002b}, without the knowledge of a Lorentzian as the best fit to the profile, the EWs were measured by numerical integration with subjectively placed limits. In our study we calculated the EW by integrating the Lorentzian profile, which exhibits extended broader wings on both sides.

It is important to note that the unresolved presence of numerous stellar photospheric lines within the profile of $\lambda$4428 does not necessarily lead to a false measurement of its equivalent width, as long as the number density and relative strengths of such lines in the adjacent continuum at both sides of the band is approximately the same as within the band. In stars of mid-A spectral type this is indeed the case, as can be seen in the R = 38,000 spectrum of the A5V star HD 194837 in Fig. \ref{fig:star46_norm}. The increased depression of the spectral intensity due to unresolved stellar lines within the band is balanced by a similar depression of the adjacent continuum due to unresolved lines present there, leaving the equivalent width of the band essentially unchanged. We carried out a detailed simulation of this process by superimposing a R = 38,000 spectrum of the unreddened star HD 194837 with a Lorentzian profile of EW = 5200 m$\angstrom$ centered at $\lambda$4428, degrading the resolution to that of our spectrograph (R = 920), and dividing the result by a similarly degraded spectrum of the unreddened A5V star HD 98058. The Lorentzian profile fit to the resulting $\lambda$4428 yielded an EW = 5213 m$\angstrom$, reproducing our imposed prior almost exactly and thus validating our measurement approach. We did not, however, correct our measured value of EW4428 in Table \ref{Tab:summery_DIBs} by this small difference, as it is well within our stated measurement uncertainty. 
 
\begin{figure}
 	\includegraphics[width=1.0\columnwidth]{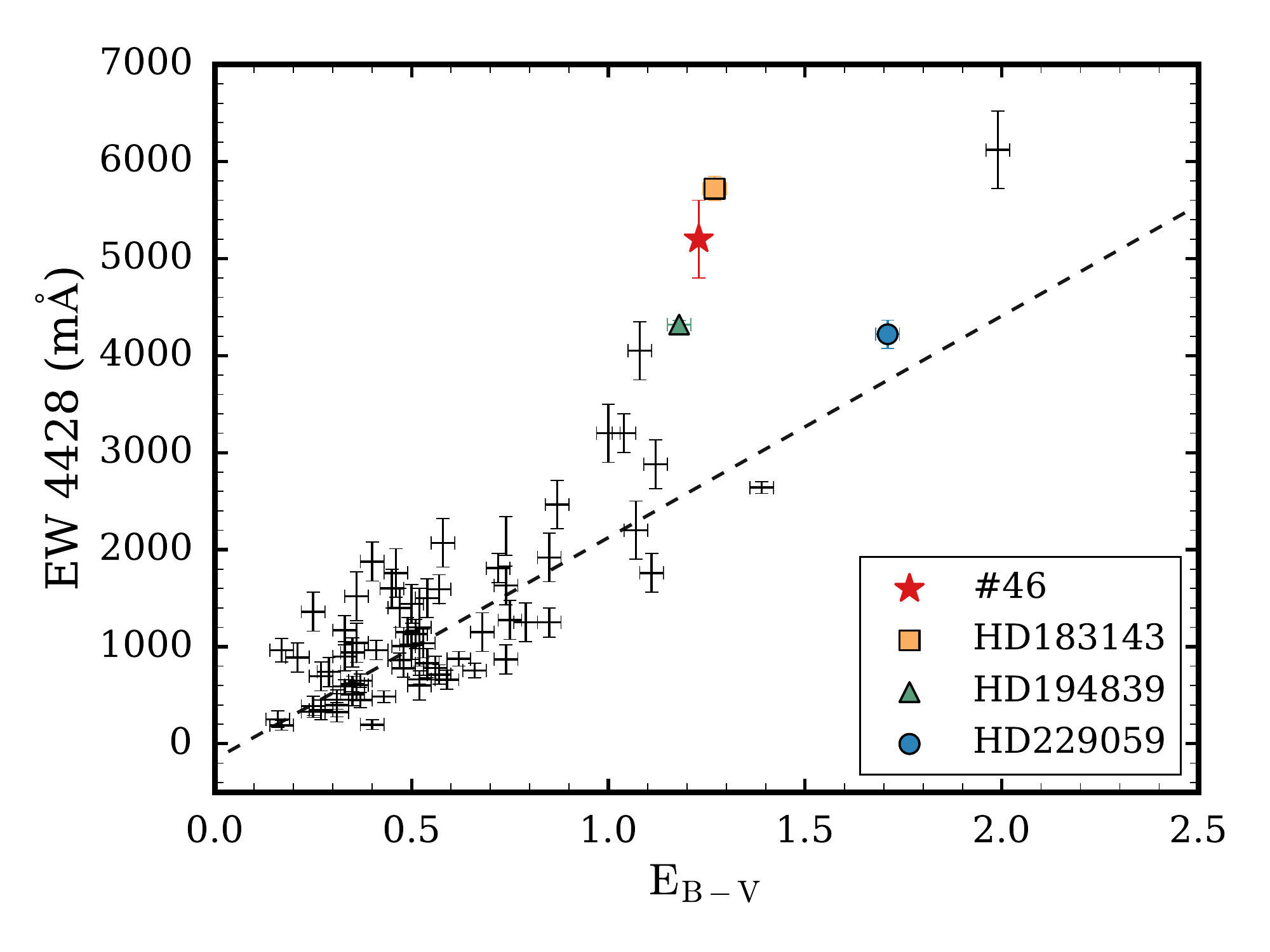}
    \caption{The $\lambda$4428 equivalent widths (EWs) measured from our study as a function of the color excess E$_{\mathrm{B-V}}$, plotted with archival data from the DIB database. The dashed line is the linear fit to the archival data. Star \#46 shows a significantly larger EW compared to the normal distribution.}
    \label{fig:DIB4428_EBV}
\end{figure}

In Fig. \ref{fig:DIB4428_EBV}, we show EW4428 as a function of E$_{\mathrm{B-V}}$ by collecting available data from the UoC DIB database (black points) together with our observations of star \#46 and our three standard stars with prominent $\lambda$4428. The dashed line is the fit to the data from the DIB database. It is not surprising that the three $\lambda$4428 standard stars, i.e. HD183143, HD 194839, and HD229059, lie above the general trend since we observed them because of their strong $\lambda$4428; but the fact that star \#46 also shows a similarly strong EW4428 is rather exceptional since, in general, DIBs are either weak or absent in nebular environments \citep{Snow1995}. A recent study of 25 significantly reddened stars by \citet{Fan2019} reported an average normalized equivalent width EW4428/E$_{\mathrm{B-V}}$ = 2006 m$\angstrom$/mag. Our result for star \#46 is EW4428/E$_{\mathrm{B-V}}$ = 4228 m$\angstrom$/mag, more than double the typical value found in the diffuse ISM. We note that there are other lines of sight with similarly high values of EW4428/E$_{\mathrm{B-V}}$, in particular, the line of sight to the fullerene planetary nebula Tc 1 \citep{Garcia-Hernandez2013, Diaz-Luis2015} with EW4428/E$_{\mathrm{B-V}}$ = 3.74, where the authors suggest a connection between DIBs and fullerenes.

\begin{figure}
 	\includegraphics[width=1.0\columnwidth]{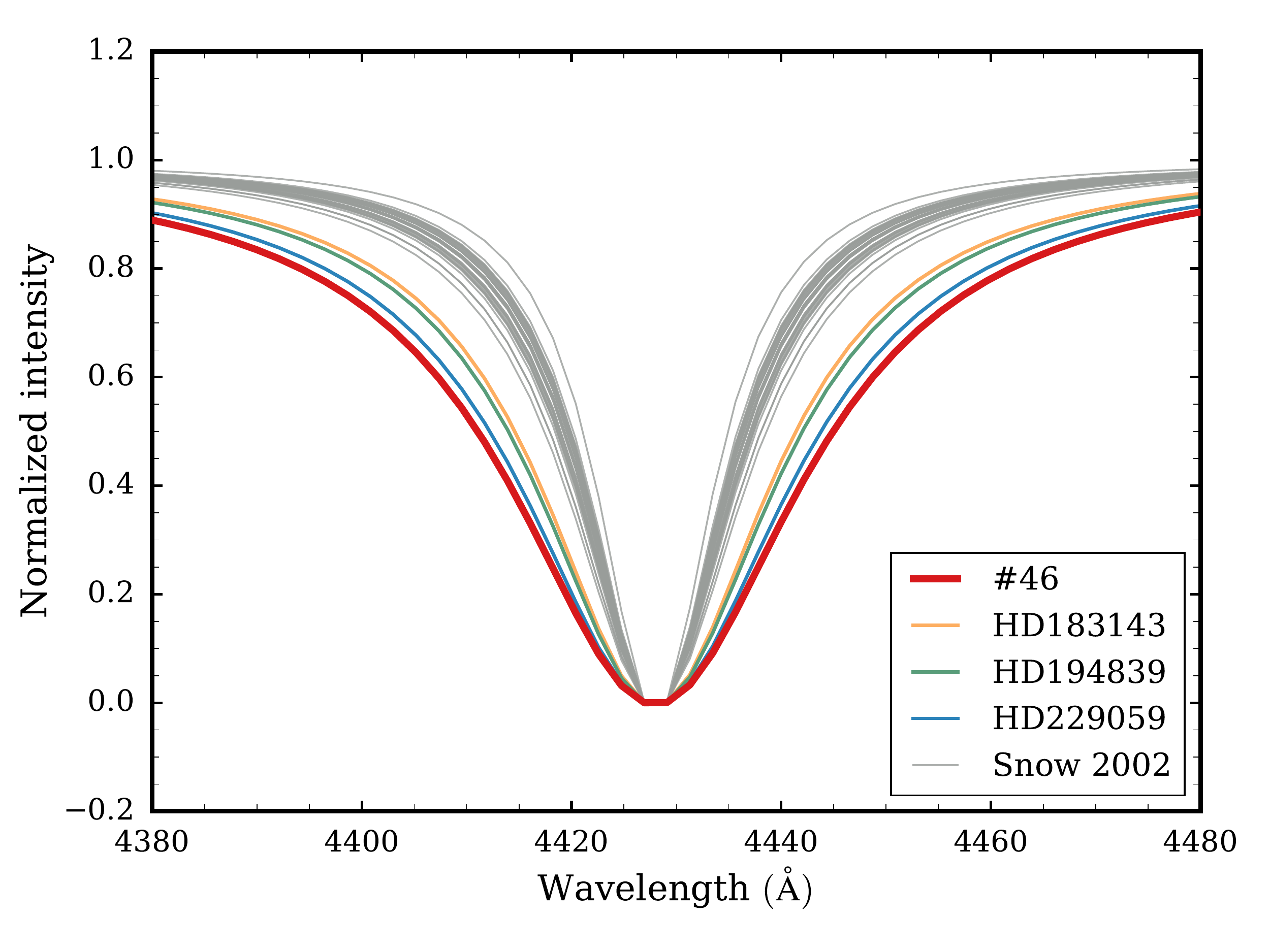}
    \caption{The fitted Lorentzian profiles of our observations together with the Cyg OB2 association measurements by \citet{Snow2002b}. Star \#46, the red profile, shows the broadest absorption feature among all of them.}
    \label{fig:width4428}
\end{figure}

The FWHM of $\lambda$4428 was measured in the fitted Lorentzian profiles for star \#46 and three $\lambda$4428 standard stars. In Fig. \ref{fig:width4428} we show these fitted profiles normalized to constant central depth in comparison with the results from \citet{Snow2002b}. The $\lambda$4428 feature found in star \#46 displays a greater FWHM compared to the $\lambda$4428 standard stars and 35 other profiles from the Cyg OB2 association measured by \citet{Snow2002b}. The FWHM of $\lambda$4428 for star \#46 was measured as 33.74 $\angstrom$, whereas the average FWHM found by \citet{Fan2019} for 25 reddened stars was 24.1 $\pm$ 5.3 $\angstrom$.

\begin{table}
	\centering
	\caption{Summary of the observed strong DIBs in IC 63 star \#46.}
	\label{Tab:summery_DIBs}
	\begin{tabular}{ccccc} 
		\hline\hline
		DIBs & FWHM & EW & EW/E$_{\mathrm{B-V}}$ & central depth \\[0.1mm]
 		& (m$\angstrom$) & (m$\angstrom$) & (m$\angstrom$/mag) & (\%)\\
        \hline
		4428 & 33740 & 5200$\pm$412 & 4228 & 12.96\\
        5780 & 4193 & 774$\pm$31 & 629 & 14.73\\
        5797 & 3100 & 233$\pm$26 & 189 &  6.00\\
        6284 & - & 1487$\pm$52 & 1209 & 17.76\\
        6614 & 2026 & 203$\pm$26 & 165 & 8.00\\
        
		\hline
%
	\end{tabular}
\end{table}	

\begin{table}
	\centering
	\caption{$\lambda$4428 equivalent width measurements in HD 183143}
	\label{Tab:HD183143_EW4428}
	\begin{tabular}{ll} 
		\hline\hline
		EW4428 & reference \\
 		(m$\angstrom$) \\
        \hline
		4920 & \citet{Greenstein1950}\\
        4240$\pm$424 & \citet{Tug1981}\\
        5700$\pm$43.3 & \citet{Hobbs2009}\\
        3570$\pm$600 & \citet{York2014}\\
        3140$\pm$276 & \citet{Krelowski2018}\\
        5724$\pm$124 & this work\\
		\hline
	\end{tabular}
\end{table}

\subsection{Other DIBs}
\begin{figure}
 	\includegraphics[width=0.5\textwidth]{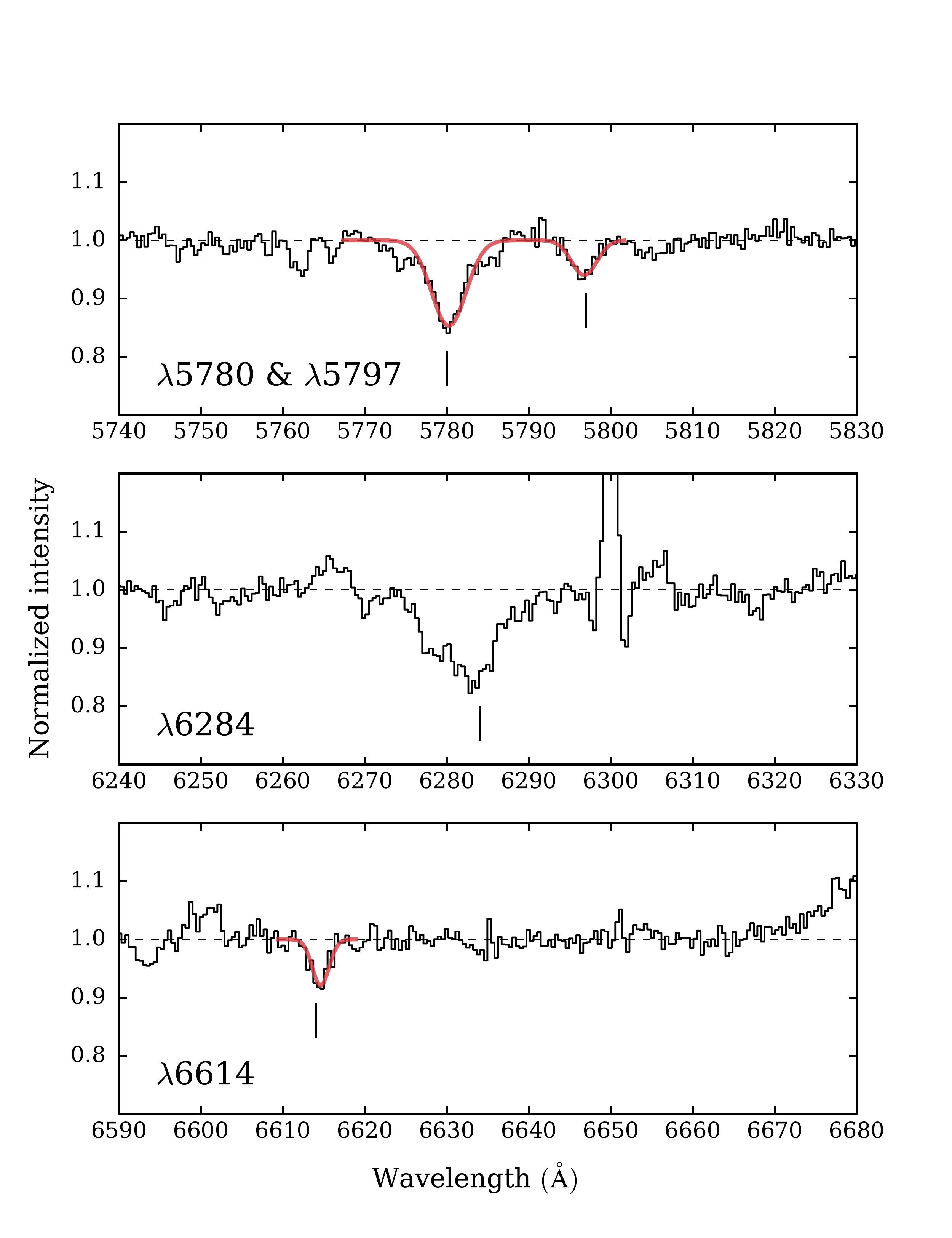}
    \caption{In addition to $\lambda$4428, several other strong DIBs are also detected in star \#46 from Keck-II/DEIMOS spectra. We estimated the EW and FWHM of $\lambda$5780, $\lambda$5797, and $\lambda$6614 using Gaussian profiles. For $\lambda$6284, its EW was measured by integrating over the absorption feature.}
    \label{fig:other_strong_DIBs}
\end{figure}

In addition to $\lambda$4428, we found several other strong DIBs in our spectrum, including $\lambda$5780, $\lambda$5797, $\lambda$6284, and $\lambda$6614 (see Fig. \ref{fig:other_strong_DIBs}). Their estimated EWs, FWHMs, and depths are listed in Table \ref{Tab:summery_DIBs}. The FWHM is measured if the profile can be fitted by a functional form. Only $\lambda$4428 is fitted by a Lorentzian, otherwise, Gaussian profiles were used for $\lambda$5780, $\lambda$5797, and $\lambda$6614. There is no sign of detection of $\lambda$4727, a so-called C$_{2}$ DIB usually associated with high column densities of C$_{2}$ molecules. This is consistent with the fact that our line of sight is observed through a PDR instead of the inner part of the molecular cloud. Other weaker DIBs present in the wavelength range covered by our spectra were not expected to be seen, given the low spectral resolution and modest S/N ratio, and they were indeed not detected.

We investigated whether our measurement of EW5780 could be compromised by contributions from unresolved Fe II lines known to be present in stars of late-B to early-A spectral type \citep{Raimond2012}. These lines are located near 5780.0 $\angstrom$, 5783.4 $\angstrom$, and 5784.0 $\angstrom$; they reach their greatest strength in late-B stars and decline in strength for both earlier and later spectral types (POLLUX Database of Stellar Spectra; http://pollux.oreme.org). We estimated the EW(Fe II 5780.0) in the spectrum of the B9 IV star HD 385 (\citealt{Raimond2012}; Fig. 2) to be not more than 20 m$\angstrom$, noting that it should be less in an A5V star like star \#46. This unresolved line, therefore, makes an insignificant contribution to our measured EW5780 for star \#46 of about 2\%, well within our stated measurement uncertainty. Our Gaussian fit to the $\lambda$5780 profile (Fig. \ref{fig:other_strong_DIBs}) excludes the contributions from the two other Fe II lines appearing in the long-wavelength wing of $\lambda$5780.

\begin{figure*}
 	\includegraphics[width=0.8\textwidth]{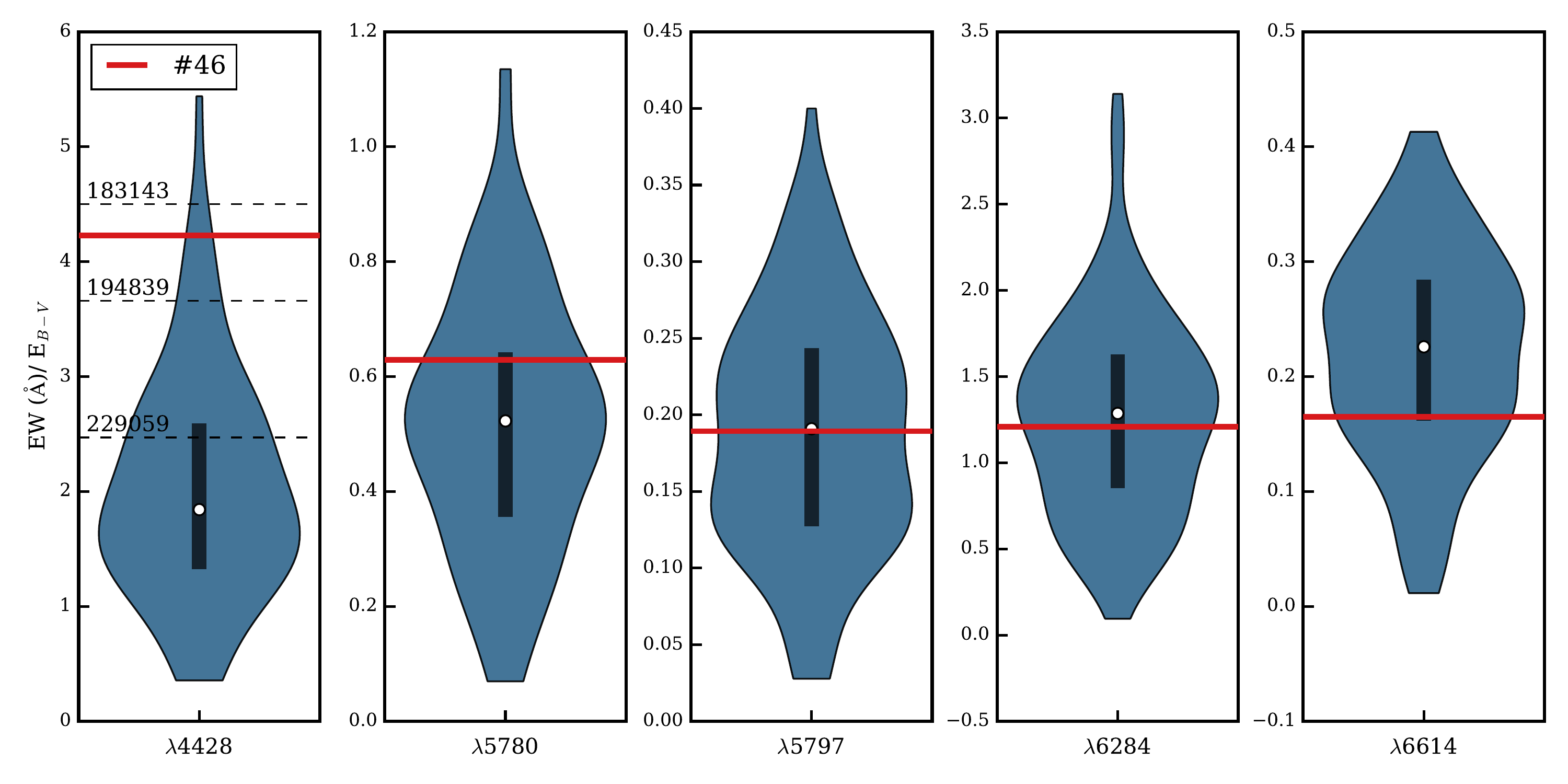}
    \caption{The distribution of the normalized EWs from the UoC DIB database. The white point in each panel indicates the median, the black bar shows the range between first and third quartiles, and the red horizontal line indicates the normalized EW of the band from star \#46. This figure shows EW4428 in star \#46 exhibits a strength more than twice that seen on average in the diffuse ISM, whereas other bands can be regarded as "normal" since they all lie within the first and third quartile of the distribution.}
    \label{fig:star46_band_strength}
\end{figure*}

We compared our EW measurement of the strong DIBs in star \#46 with the data retrieved from the UoC DIB database. In Fig. \ref{fig:star46_band_strength} we show the distribution of the EW normalized to the color excess for 98 stars with E$_{\mathrm{B-V}}$ $>$ 0.2. Stars with E$_{\mathrm{B-V}}$ $<$ 0.2 are not included due to the larger uncertainties in their measurements. The white point in each panel indicates the median, the black bar shows the range between the first and third quartile, and the red horizontal line indicates the normalized EW of the band from star \#46. The normalized $\lambda$4428 EWs of the three standard stars are shown in dashed lines in the leftmost panel. Taken all together, we clearly see that the EW of $\lambda$4428 in star \#46 is exceptionally large while the other bands lie between the first and third quartile. This shows, again, the uniqueness of this line of sight towards star \#46, containing an exceptionally strong $\lambda$4428 feature.

\subsection{Search for RR emission features}
\begin{figure}
 	\includegraphics[width=1.0\columnwidth]{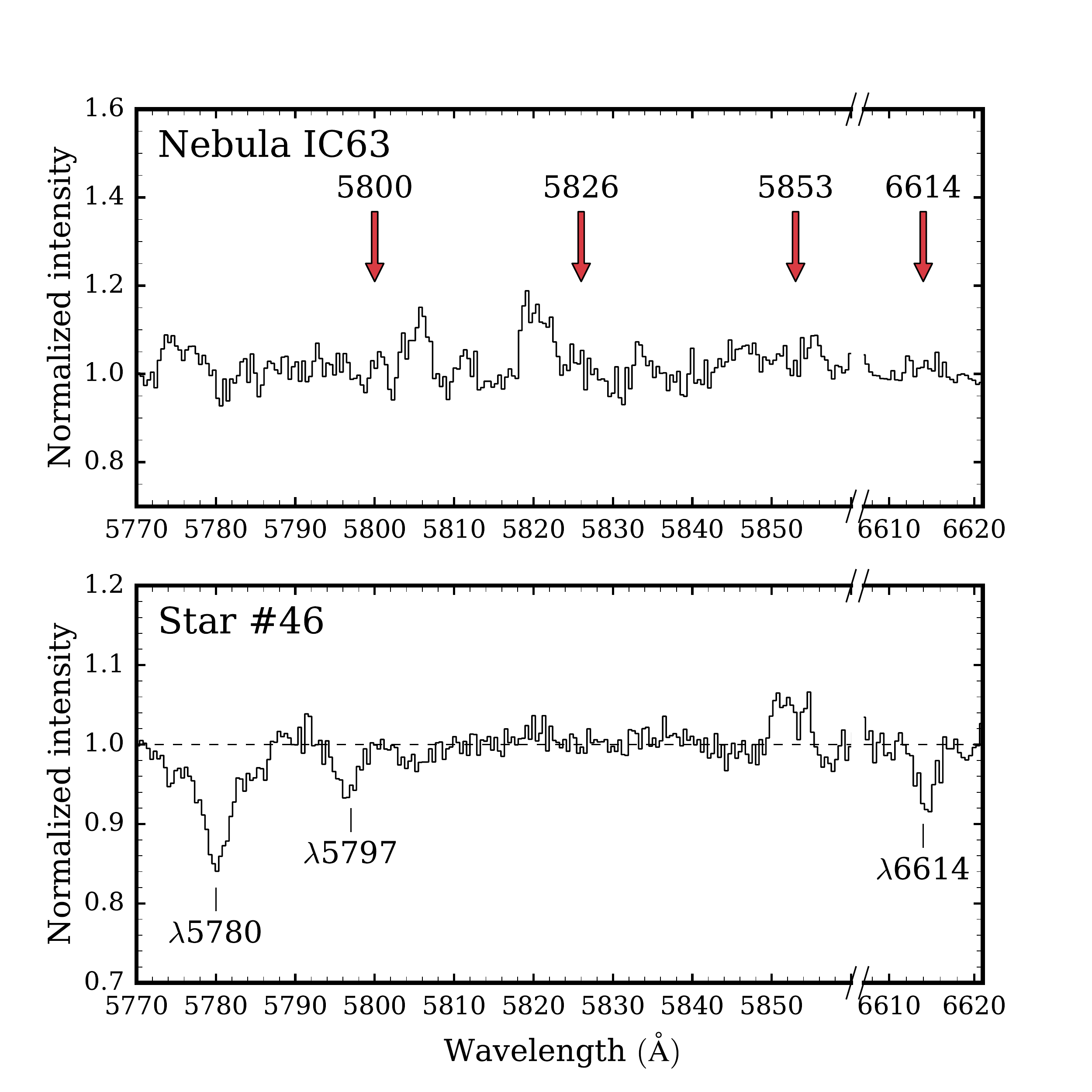}
    \caption{The Keck-II/DEIMOS spectra of the surrounding nebula and star \#46 are shown with identical wavelength coverage. The four red arrows indicate where the sharp RR-features were expected to appear in the spectrum of IC 63, if present.}
    \label{fig:5800_emission}
\end{figure}

With both the stellar and nebular spectra in hand, we used the line of sight of star \#46 to search for the nebular sharp emission features which were hypothesized to be the counterpart of DIBs. As demonstrated in Fig. \ref{fig:5800_emission} we found no sign of such emission features in the spectral range where DIBs were clearly detected in the stellar spectrum. If present with the same relative intensity compared to the ERE continuum as observed in the Red Rectangle, the sharp RR features should rise to a level near 2.0 on the normalized intensity scale shown. This is an argument against the claim that the sharp RR-features and certain DIBs share the same carriers.

\section{Discussion}
\label{discussion}
\subsection{Connection between DIBs and ERE}
In Sec. \ref{DIB_ERE_connection}, we listed a number of plausible arguments which made a close connection between the respective carriers of DIBs and ERE appear likely. Nevertheless, our detection of strong DIBs in the IC 63 ERE filament via the spectrum of star \#46 was unexpected, given that, normally, DIBs are weak or absent in dense or strongly irradiated clouds (\citealt{Snow1995}; \citealt{Jenniskens1994}). The fact that our line of sight probes a region with the most intense ERE in IC 63 and simultaneously shows evidence of strong DIBs, in particular an exceptionally strong and broad $\lambda$4428, makes it even more likely that either the carriers of ERE and these DIBs are the same or at least very closely related molecules that are produced in abundance by the processes occurring in the PDR. Our findings are consistent with the suggestion that DIB carriers survive in the hostile environment of the IC 63 PDR, because the ERE process via recurrent fluorescence serves as an effective cooling mechanism for the common carriers. ERE in the general diffuse ISM \citep{Gordon1998} likely serves the same role for DIB carriers in the diffuse ISM. Furthermore, the observed coexistence of ERE and some DIB carriers strongly supports the Poincar\'e fluorescence model \citep{Leger1988} for the ERE,  which attributes the origin of ERE to small carbonaceous molecular ions with 18--36 carbon atoms. It has been a consensus within the DIB field that the carriers of DIBs are molecules or molecular ions in the same general size range. On the other hand, our findings do not favor other suggested ERE carrier models such as hydrogenated amorphous carbon \citep{Duley1990}, silicon nanoparticles \citep{Witt1998}, or graphene oxide \citep{Sarre2019}.

The line of sight to star \#46 penetrates the IC 63 PDR at a location immediately behind the hydrogen ionization front and in front of regions where rotationally excited molecular hydrogen (Fig. \ref{fig:star46_ERE}) and vibrationally excited neutral and ionized PAHs are observed \citep{Lai2017}. PDR models show that our line of sight most likely penetrates an environment where molecular hydrogen has largely been dissociated by stellar radiation from $\gamma$ Cas in the energy range 11.2 eV--13.6 eV, i.e. it is a medium dominated by atomic hydrogen. With molecular hydrogen eliminated as a competing opacity source, ERE carriers are readily excited by photons with E < 13.6 eV \citep{Witt2006} and recurrent fluorescence provides a path to rapid cooling of the excited molecules. This provides a mechanism for extending the lifetime of otherwise marginally stable species. The abundance of these particles is most likely the result of a balance of topdown production by radiative torque disruption of larger grains \citep{Hoang2018} and photo-processing of nanoparticles \citep{Jones2016} on one hand and photo-dissociation by FUV photons \citep{Allain1996} and dissociative recombination on the other.

\subsection{A broader $\lambda$4428 profile}
The Lorentzian profile fit to the $\lambda$4428 implies that the broadening of the band is caused by natural line broadening. Its large width stems from an exceptionally short lifetime of the upper level of the transition, which is most likely depopulated by internal conversion \citep{Snow2002b}. The notion of internal conversion after photon absorption is also supported by the fact that no substructure was found in $\lambda$4428 \citep{Snow2002a}. The characteristic width of the profile is proportional to the transition rate, i.e. the inverse of the upper state lifetime governed by Heisenberg's uncertainty principle. \citet{Snow2002b} showed that the upper state lifetime of the transition is of the order of $\sim$10$^{-13}$ s, by assuming an average $\lambda$4428 profile FWHM of 17.25 $\angstrom$. The only process which can account for such a rapid transition is internal conversion, a radiationless transition that redistributes the excitation energy into numerous internal energy states without any photon relaxation. Even though internal conversion may be responsible for such transitions, the observed wide variation of the width of the profile seen in Figure \ref{fig:width4428} is still unexplained; it is our objective to address a possible mechanism for the variation in the broadening of the absorption feature in different lines of sight. We note that Doppler broadening can not explain the width of $\lambda$4428 because such a width could only be achieved by a velocity dispersion of $\sim$2300 km s$^{-1}$, which is unlikely, considering the gas kinematics in IC 63.

Thus, a broader Lorentzian profile must be the consequence of a shorter lifetime of the upper state during the internal conversion transition. A mechanism that is able to reduce the transitional lifetime is the key to address such a broadening effect. \citet{Amirav1987} showed that the electronic relaxation timescale can be altered by the effect of Coriolis rotational-vibrational interaction on the redistribution of the intramolecular vibrational states. The increase of the rotational energy of molecules expected in a high-radiation environment leads to states that are vibrationally coupled, making the energy level distribution denser and more continuum-like as opposed to the discrete distribution found when ignoring the rotation effect of the molecule \citep{Amirav1988}. Thus, the width of the observed profile becomes wider as a result of the increase in the intramolecular transition probability. The caveat for such argument, however, is to ensure that the molecules were collectively retained in a rotationally excited equilibrium state, represented by a stable rotation temperature, for time periods longer than the average interval between photon absorption.

It is known that the characteristic timescale for absorbing a UV-photon is comparable to the emission of a rotational photon in the diffuse interstellar medium, whereas in a refection nebula such as IC 63 the timescale for photon absorption is much shorter than the timescale for rotational damping (see Table 1. in \citealt{Silsbee2011}). We applied equation (12) in \citet{Yard2010} to calculate the damping rate by rotational emission for a change in rotational quantum number of $\Delta J = \pm 1$. If assuming N$\mathrm{_{c}}$ = 50 carbon atoms, this leads to a damping timescale of the order of 5 days, a timescale much larger than that for photon absorption followed by mid-IR emissions, which are of the order of hours or seconds, respectively. We assume, in accordance with \citet{Yard2010} and \citet{Hoang2018}, that photon absorption followed by mid-IR vibrational emissions are the dominant contributors to establishing the rotational temperature of the molecules.

The molecules, therefore, are able to build up an angular momentum distribution that peaks at a higher J value, and at the same time remain in an equilibrium stationary state \citep{Mulas1998}. In other words, for an individual molecule in a nebula such as IC 63, its rotational temperature is static, making it an ideal probe for the surrounding environment. In the case of IC 63, the anisotropic radiation field contributes radiative torque, expedites the spinning of the molecules and builds up a stationary state that has a higher rotational energy distribution \citep{Hoang2018}. This leads to the conclusion that the excessive broadening of $\lambda$4428 in the spectrum of star \#46 is a consequence of the shorter timescale in the upper state due to stronger Coriolis rotational-vibrational interactions. The wide range of values in the FWHM of $\lambda$4428 illustrated in Figure \ref{fig:width4428} can then be understood in terms of different rotational temperatures of the carrier molecules in different lines of sight, which is a consequence of differences in the local radiation field.

\subsection{EW5797/EW5780 as a probe of UV radiation field strength}
The strongly variable EW5797/EW5780 ratio was suggested as a probe for the density of the UV radiation field (\citealt{Vos2011}). \citet{Krelowski1992} showed that there is a dichotomy in cloud families: $\zeta$-type clouds --- UV shielded sightlines passing through dense clouds with higher EW5797/EW5780 ratios and $\sigma$-type clouds --- sightlines through the skin of clouds that are non-shielded from the UV radiation, with lower EW5797/EW5780 ratios. The separation of the two types of clouds was defined by the ratio of central depths of $\lambda$5797 and $\lambda$5780 (A5797/A5780 = 0.4), which corresponds to a ratio of EW5797/EW5780 $\sim$ 0.3 \citep{Kos2013}. However, the existence of intermediate sightlines leaves the transition of the two cloud types rather smooth, making it difficult to make a sharp distinction between $\sigma$ and $\zeta$-type clouds \citep{Krelowski2019}. 

The carriers of $\lambda$5780 and $\lambda$5797 appear to react very differently to UV radiation. The ionization potential of the $\lambda$5780 carriers are larger, so a harder radiation field is required to excite the carriers. The carriers of  $\lambda$5797, on the other hand, can be excited by the photons with less energy deep inside the cloud and are largely destroyed in HII regions \citep{Cami1997, Sonnentrucker1997}. We found the EW5797/EW5780 ratio towards star \#46 to be 0.30, a value which is representative of the average ISM with low UV radiation field intensities. It lies near the boundary of $\zeta$/$\sigma$-type clouds (the aforementioned EW5797/EW5780 $\sim 0.3$; \citealt{Vos2011}). Also, with measured EWs of $\lambda$5780 and $\lambda$5797 at 774 m$\angstrom$ and 233 m$\angstrom$, respectively, and E$_{\mathrm{B-V}}$= 1.23, both DIBs closely follow the correlations with E$_{\mathrm{B-V}}$ for Galactic or extragalactic sightlines \citep{Monreal-Ibero2018, Weselak2019}. Hence, caution is needed when attempting to use EW5797/EW5780 as a direct probe of the ISRF. In fact, while it is certain that a higher EW5797/EW5780 ratio ($\sim$ 1) suggests a lower ISRF, a wide density range of the radiation field, from G$_{0}$ = 1--20 as illustrated by \citet{Vos2011}, or as an extreme case in IC 63 where G$_{0} \gtrsim 150$, may lead to similar EW5797/EW5780, especially for $\sigma$-type clouds. 

Alternatively, $\lambda$5780 and $\lambda$5797 show a robust correlation with atomic and molecular hydrogen. $\lambda$5780 is well correlated with HI but insensitive to H$_{2}$, whereas $\lambda$5797 shows a positive relation with H$_{2}$ \citep{Herbig1993, Welty2014, Lan2015, Weselak2019}. Together, this leads to the view that EW5797/EW5780 correlates well with the fraction of molecular hydrogen, f$_{\mathrm{H_{2}}}$ (see Fig. 5A in \citealt{Fan2017}), and it suggests that EW5797/EW5780 is primarily affected by the relative abundance of atomic and molecular hydrogen, but only indirectly affected by the UV radiation which can dissociate the molecular hydrogen. With the correlation between f$_{\mathrm{H_{2}}}$ and EW5797/EW5780 \citep{Fan2017}, the inferred f$_{\mathrm{H_{2}}}$ for star \#46 is in the range 0.1--0.2, which is consistent with a low f$_{\mathrm{H_{2}}}$ under a high radiation field described in \citet{Vos2011} (See Fig.20). So, atomic hydrogen is predominant in the line of sight towards star \#46, where intense photo-processing dissociates molecular hydrogen into atomic form. 

Both the strong radiation field (G$_{0}\gtrsim$150) and the high gas density, n(H$_{\mathrm{2}}$) = (5 $\pm$ 2) $\times$ 10$^{4}$ cm$^{-3}$ \citep{Jansen1995}, in the PDR of IC 63 play important roles in regulating the ratio of EW5797/EW5780. \citet{Pilleri2012} pointed out that the carbonaceous molecules in the PDR are likely produced by photo-evaporation of very small grains, and \citet{Jones2016} proposed that DIB carriers were likely formed by a top-down fragmentation process from a-C(:H) nanoparticles. These ideas seem to be consistent with the scenario in IC 63 where the intense UV radiation field drives the photo-fragmentation process, leading to an increase in the density of DIB carriers. Hence, a possible explanation for having a typical rather than an extremely low value of EW5797/EW5780 towards star \#46 is the increase of the abundance of $\lambda$5797 carriers that balances the favored $\lambda$5780 band in the low f$_{\mathrm{H_{2}}}$ environment. To conclude, the ratio of EW5797/EW5780 is not a probe of the UV radiation field in an environment with a very high gas density, where dissociation and recombination of hydrogen are balanced and maintained at levels similar to those found in the low-density ISM.

\subsection{Absence of the sharp RR-emission features}
The long-slit observation to star \#46 allowed us to conduct a direct test of whether DIBs and the RR-features are related by collecting the stellar and nebular spectra simultaneously from the same line of sight. In contrast, previous studies that suggested the RR-features are the counterpart of the DIBs came from observations of the RR-features in the absence of simultaneous DIB detections \citep{Scarrott1992, Sarre1995, Duley1998}; the fact that no corresponding DIBs were found through the RR nebula cast doubt on the hypothesis of common carriers \citep{Hobbs2004}. Again, if this hypothesis holds true, given that the illuminating star, $\gamma$ Cas, provides a strong radiation field that can easily excite DIB carriers and thus produce the RR-features, the emission features should be detectable, with intensity as strong as the underlying continuum \citep{Schmidt1991}. And yet, our data show no sign of the RR-features even when the corresponding DIBs ($\lambda$5780, $\lambda$5797 and $\lambda$6614) are clearly seen (see Fig. \ref{fig:5800_emission}). Thus, we regard our result as evidence that the RR-features and the nearby DIBs are likely not related. Finally, we note that the sharp band features in the RR are a unique phenomenon, which so far has only been detected in RR and the circumstellar envelope of the R Coronae Borealis star V854 Centauri \citep{Rao1993, Oostrum2018}. We note that no ERE was detected in the case of V854 Centauri.

\section{Summary}
\label{summary}
The goal of this paper is to try to test the hypothesis that the carriers of DIBs and ERE are the same. The line of sight towards star \#46 presents a perfect geometry to test such an hypothesis by enabling observations of both the strongest DIBs and ERE simultaneously. Our data and analysis led to the following conclusions:

1. We observed 5 typically strong DIBs along the line of sight towards star \#46 seen through the reflection nebula IC 63, where the most intense ERE is detected. The presence of these DIBs along this sightline, with the normalized strength of $\lambda$4428 stronger by more than a factor of two than found in typical interstellar sightlines, is indeed consistent with the hypothesis that the carriers of these DIBs and ERE are identical. We have shown that carriers of both phenomena coexist in a PDR environment normally viewed as hostile to molecules that are considered likely DIB carriers.

2. We argued that the 5 detected DIBs share similar characteristics as ERE carriers in terms of their preference for an environment dominated by atomic hydrogen. We suggest that ERE, produced by recurrent fluorescence, enabled by rapid inverse internal conversion, provides effective radiative cooling that allows the common carrier molecules to survive in regions of space dominated by photons with E $<$ 13.6 eV.

3. Compared with DIB standard stars that all show prominent $\lambda$4428 bands with Lorentzian profiles, the spectrum of star \#46 exhibits the widest $\lambda$4428 profile. We attributed the excessive broadening of $\lambda$4428 in star \#46 to a shorter lifetime in the upper state of the transition. We suggest that Coriolis rotational-vibrational interaction caused by the higher rotational temperature in the IC 63 PDR is responsible for increasing the transition probability for internal conversion. 

4. We found EW5797/EW5780 = 0.30 for the line of sight towards star \#46, a value comparable to sightlines through the diffuse ISM with typically low interstellar radiation field densities. We conclude that the ratio of EW5797/EW5780 is not a sensitive indicator of the density of the UV radiation field. Instead, it appears to measure the fractional molecular hydrogen abundance along the line of sight, which depends on both the densities of gas and UV radiation.

5. We were able to test the hypothesis that the carriers of the sharp RR-features and the nearby DIBs are the same by enabling observation of both the stellar and nebular spectra towards star \#46. We did not detect the sharp RR-features even though DIBs are clearly detected. This result suggests that the RR-features and the nearby DIBs do not share common carriers.

\section*{Acknowledgements}
The authors thank the anonymous referee for the careful reading and useful feedback. These results made use of the Discovery Channel Telescope at Lowell Observatory. Lowell is a private, non-profit institution dedicated to astrophysical research and public appreciation of astronomy and operates the DCT in partnership with Boston University, the University of Maryland, the University of Toledo, Northern Arizona University and Yale University. In addition, some of the data presented herein were obtained at the W. M. Keck Observatory, which is operated as a scientific partnership among the California Institute of Technology, the University of California and the National Aeronautics and Space Administration. The Observatory was made possible by the generous financial support of the W. M. Keck Foundation. The authors wish to recognize and acknowledge the very significant cultural role and reverence that the summit of Maunakea has always had within the indigenous Hawaiian community.  We are most fortunate to have the opportunity to conduct observations from this mountain. We also acknowledge the UoC database for providing us the rich dataset to make this study possible. 

TL gratefully acknowledges support for this project from the University of Toledo Graduate Fellowship. 

JC acknowledges support from an NSERC Discovery Grant and a Western University SERB Accelerator Award.

We thank Teznie Pugh, Jason Sanborn, and Heidi Larson at Lowell Observatory for technical support during our observation. Finally we thank Prof. Aviv Amirav for helpful communications and useful suggestions. 





\bibliographystyle{mnras}
\bibliography{DIBs} 




\appendix




\bsp	
\label{lastpage}
\end{document}